# Realizing gapped surface states in magnetic topological insulator MnBi$_{2-x}$Sb$_x$Te$_4$

Wonhee Ko[1], Marek Kolmer[1], Jiaqiang Yan[2], Anh D. Pham[1], Mingming Fu[1,3], Felix Lüpke[1,4], Satoshi Okamoto[2], Zheng Gai[1], Panchapakesan Ganesh[1], An-Ping Li[1*]

[1]Center for Nanophase Materials Sciences, Oak Ridge National Laboratory, Oak Ridge, Tennessee 37831, USA

[2]Materials Science and Technology Division, Oak Ridge National Laboratory, Oak Ridge, Tennessee 37831, USA

[3]Fujian Provincial Key Laboratory of Semiconductors and Applications, Collaborative Innovation Center for Optoelectronic Semiconductors and Efficient Devices, Department of Physics, Xiamen University, Xiamen, Fujian Province 361005, P.R. China

[4]Department of Materials Science and Engineering, University of Tennessee, Knoxville, Tennessee 37916, USA

[*]Correspondence to apli@ornl.gov

**The interplay between magnetism and non-trivial topology in magnetic topological insulators (MTI) is expected to give rise to a variety of exotic topological quantum phenomena, such as the quantum anomalous Hall (QAH) effect and the topological axion states [1,2]. A key to assessing these novel properties is to tune the Fermi level in the exchange gap of the Dirac surface band. MnBi$_2$Te$_4$ possesses non-trivial band topology with intrinsic antiferromagnetic (AFM) state that can enable all of these quantum states [3-7], however,**

**highly electron-doped nature of the MnBi$_2$Te$_4$ crystals obstructs the exhibition of the gapped topological surface states [8-10]. Here, we tailor the material through Sb-substitution [11,12] to reveal the gapped surface states in MnBi$_{2-x}$Sb$_x$Te$_4$ (MBST). By shifting the Fermi level into the bulk band gap of MBST, we access the surface states and show a band gap of 50 meV at the Dirac point from quasi-particle interference (QPI) measured by scanning tunneling microscopy/spectroscopy (STM/STS). Surface-dominant conduction is confirmed below the Néel temperature through transport spectroscopy measured by multiprobe STM. The surface band gap is robust against out-of-plane magnetic field despite the promotion of field-induced ferromagnetism. The realization of bulk-insulating MTI with the large exchange gap offers a promising platform for exploring emergent topological phenomena.**

MnBi$_2$Te$_4$ is a MTI which has inverted band structure from large spin-orbit coupling and magnetically induced gap between electronic energy bands [3-5]. The crystal has layered structure with septuple layers (SL) of Te-Bi-Te-Mn-Te-Bi-Te (inset of Fig 1a). Below the Néel temperature, magnetic moments of Mn atoms form A-type AFM ordering where they align ferromagnetically in the same layer and antiferromagnetically between different layers [6,7]. The topological nature of the compound was confirmed by observing the topological surface states with angle-resolved photoemission spectroscopy (ARPES) [5,8-10,12-14]. However, the abundant bulk carriers and their spatial fluctuations in the highly electron-doped MnBi$_2$Te$_4$ crystals hinder the realization of predicted topological phenomena. For example, previous ARPES studies reported both the existence [5,13,14] and the absence of an exchange gap in the topological surface states [8-10]. Thus, to resolve the fine structures of the band around the Dirac point, it is necessary to control the carrier density and use local spectroscopic techniques like STM/STS [15-17].

To compensate the electron carriers in the bulk, thin films of MnBi$_2$Te$_4$ were adapted to reduce bulk carriers and back gating was used to shift Fermi level inside the gap. Based on this approach, the transport measurements indeed demonstrated some characteristics of MTI such as QAH effect [18] and axion insulator states [19]. However, geometrical effects like interaction between the top and bottom surfaces can alter the topology of the system [20], and impurities and defects introduced during the exfoliation or the epitaxial growth can limit the film quality [21]. On the other hand, an approach based on bulk single crystals and *in situ* characterization can significantly overcome these issues to offer a system with higher quality. In particular, substituting Bi with Sb provides means to introduce hole-doping [11,12], and the change in the composition of MBST enables to alter the topological phases as well (Fig. 1a). For example, topological phase transitions are expected in increasing Sb composition *x*, where the size of the (inverted) band gap of MnBi$_2$Te$_4$ reduces and changes its sign to become a topologically-trivial normal insulator (NI) like MnSb$_2$Te$_4$ (Fig. S4) [12]. In addition, magnetic structure of MBST can change with composition [22]. Topological phase transitions are predicted for different magnetic structures such as topological insulator (TI) or AFM TI for AFM structure and ferromagnetic (FM) insulator or Weyl semimetal (WSM) for FM structure [3,4]. Therefore, shifting the Fermi level into the exchange gap through tailoring the MBST composition offers a key not just to accessing the surface band structures but also to unlocking many of these emergent quantum properties of intrinsic MTIs.

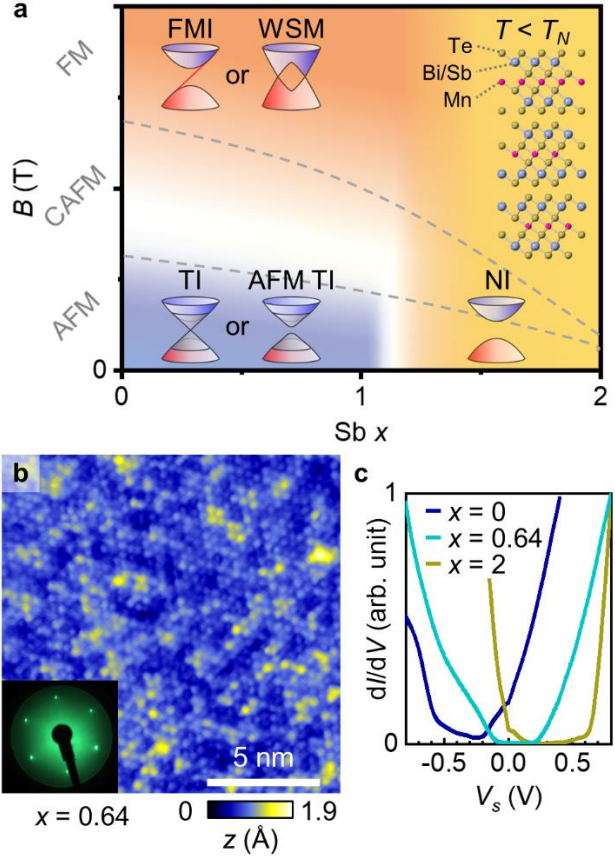

**Fig. 1. Topological phase diagram of MBST and STM/STS measurements. a.** Schematic phase diagram of MBST with respect to Sb composition $x$ and magnetic field $B$. Inset on the right shows the MBST lattice structure. **b.** STM image of MBST with $x = 0.64$ ($V_s = -0.15$ V, $I = 100$ pA). Inset shows the LEED image of the same sample at 60 eV. **c.** $dI/dV$ spectra for MBST crystals with different compositions. Each curve represents the average of about 1000 spectra taken over the area of $30 \times 30$ nm$^2$ (set point with a tunneling resistance $R_T = 2$, 15, and 0.4 GΩ for $x = 0$, 0.64, and 2, respectively).

For systematic study of the topological phases, we chose three MBST compositions of $x = 0$, 0.64, and 2. Here, $x = 0$ and 2 represent the exemplary cases of topological and trivial compounds, respectively, and $x = 0.64$ represents the bulk-insulating case that is expected to be topological (Fig. 1a). Figure 1b shows a typical topographic image of the cleaved surface of MBST, which displays triangular lattice of the topmost Te layer. The long-range crystallinity of the surface is confirmed by low energy electron diffraction (LEED) displaying hexagonal patterns (inset of Fig. 1b). The effect of Sb substitution is examined by taking $dI/dV$ spectra at $T = 4.6$ K, which is well below the Néel temperature $T_N \approx 20$ K [6,7,11]. As expected, hole-doping from Sb shifts the spectra toward the higher bias with increasing $x$. The spectrum for $x = 0$ shows V-shape curves of typical TIs [23-25], while the spectrum for $x = 2$ shows U-shape curves with a flat zero amplitude region which is typical for trivial NIs [26]. Interestingly, $x = 0.64$ compound displays a flat bottom around the Fermi level as a bulk-insulating compound, enabling access to the surface states.

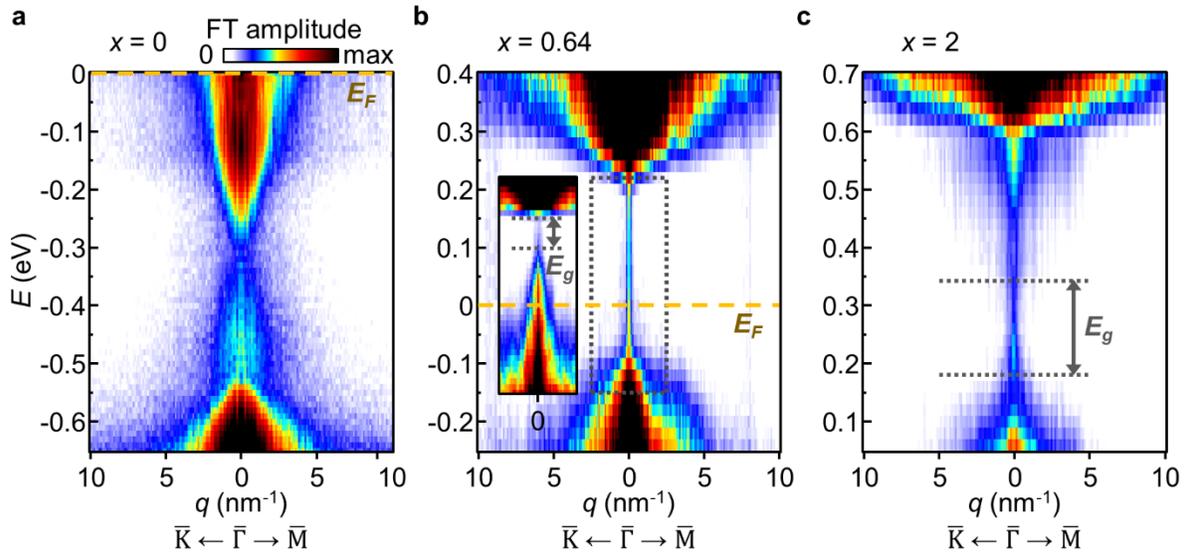

**Fig. 2. Electronic band structures of MBST mapped by QPI. a-c.** Line cuts of QPI along $\bar{\Gamma}$-$\bar{K}$ and $\bar{\Gamma}$-$\bar{M}$ directions stacked in energy for MBST with $x = 0$, 0.64, and 2, respectively (set point $R_T$ = 0.81, 1.67, and 0.4 GΩ for panel **a-c**, respectively). The energy of the states $E$ relative to the Fermi energy $E_F$ is calculated as $E - E_F = eV_s$. Inset in (**b**) shows the QPI taken with the STM tip closer to the surface (set point $R_T = 1$ GΩ).

To evaluate the surface band gap, we employed the QPI measurement to reveal surface electronic structures [27] for different MBST compositions (see movies in Supplementary Information). Figure 2 displays the line cuts of the QPI maps along $\bar{\Gamma}$-$\bar{K}$ and $\bar{\Gamma}$-$\bar{M}$ directions in the Brillouin zone that are stacked vertically in energy. The $x = 0$ and $x = 2$ compounds show distinct features originating from their topologies. The QPI map of $x = 0$ compound (Fig. 2a) shows strong intensity of dispersing bands at $E < -0.53$ eV and $E > -0.25$ eV which presumably correspond to the bulk valence and conduction bands, respectively. Inside the bulk band gap, there is another dispersing band with weaker intensity, indicative of surface states [23,25]. However, a significant background signal from bulk carriers makes it hard to determine the surface band gap around the Dirac point at $E = -0.3$ eV. In contrast, the $x = 2$ compound (Fig. 2c) shows a large bulk band gap at $0.18$ eV $< E < 0.34$ eV without any dispersing surface bands inside as expected for trivial NIs. Interestingly, the $x = 0.64$ compound (Fig. 2b) exhibits the bulk band dispersion similar to the $x = 0$ compound at $E < -0.08$ eV and $E > 0.2$ eV, with more pronounced bulk band gap due to the large reduction of the bulk carrier density. Still, when compared with the $x = 2$, the $x = 0.64$ compound shows finite d$I$/d$V$ signal inside the bulk band gap region, especially when the STM tip is brought closer to the surface (Fig. S1). QPI map at a close tip-sample distance reveals additional dispersing bands inside the bulk band gap (inset of Fig. 2b). Two almost linearly dispersing bands form a Dirac-cone-like structure with strongly reduced intensity at the crossing point. These emerging bands do not originate from external effects, such as the tip-induced band bending of the bulk bands [26,28,29], as verified by various tip height set points (Fig. S2). At the Dirac point, the surface band shows a clear gap of $E_g = 50$ meV (inset of Fig. 2b and Fig. S1). The surface band gap with size of about 40 meV on the (111)-projected surface of the $x = 0.64$ compound is reproduced by our density functional theory (DFT) calculations, which show the existence of the non-trivial topological phase with A-type AFM coupling (Fig. S4) with a massive Dirac cone (Fig. S5). The gap is due to uncompensated magnetic moments and breaking of translational symmetry on the (111) surface, while the projected surface states on the (110) surface still retain the gapless Dirac dispersion due to the combined time-reversal and translational symmetries (Fig. S6). The good agreement between STM observations and DFT calculations allows us to conclude that the bulk band of the $x = 0.64$ compound is inverted and the topological surface states on the (111) surface are gapped [3-5,30].

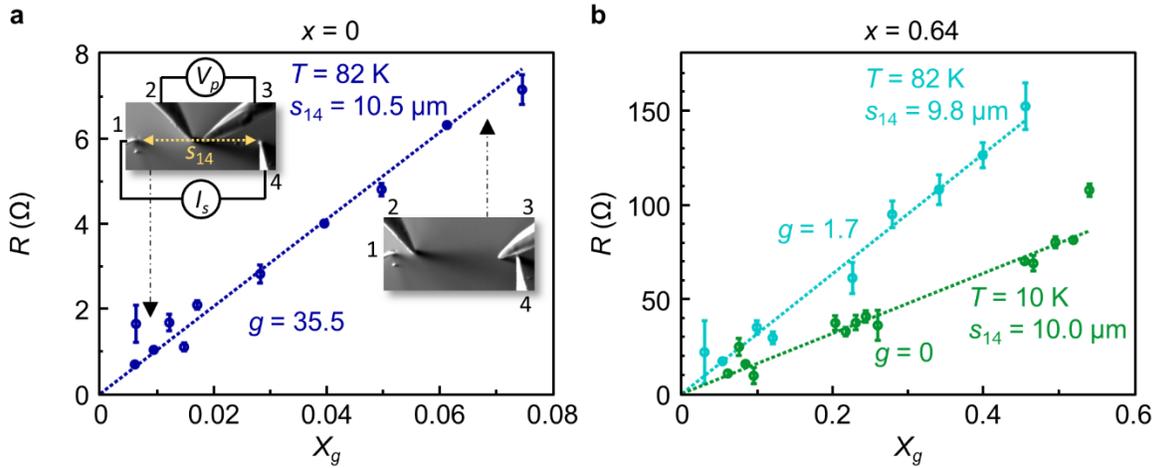

**Fig. 3. Bulk and surface conduction differentiated by variable probe-spacing transport spectroscopy. a.** The $x = 0$ compound shows bulk-dominant conduction. Linear fitting of $R$-$X_g$ graph taken at $T = 82$ K (dotted line) gives rise to $g = 35.5$ with $\rho_{3D} = 3.04$ mΩ·cm and $\rho_{2D} = 102$ Ω. Insets are the SEM images with the schematic of variable probe-spacing measurement performed by four-probe STM (SEM image size 24.5 μm × 12 μm). **b.** The $x = 0.64$ compound shows surface-dominant conduction with a bulk-insulating behavior. Linear fittings of the graphs

(dotted lines) give rise to $g = 1.7$, $\rho_{3D} = 182$ mΩ·cm and $\rho_{2D} = 318$ Ω at $T = 82$ K, and $g = 0$, $\rho_{3D} > 1590$ mΩ·cm and $\rho_{2D} = 159$ Ω at $T = 10$ K.

To further confirm the surface nature of the electronic states at the Fermi level in the bulk-insulating MBST, we measured surface transport with a multiprobe STM [31]. We employed a variable probe-spacing transport spectroscopy method based on the multiprobe STM, which was developed by us to efficiently differentiate the bulk and surface conductance [32-34]. Briefly, four-probe resistance $R$ is measured while varying probe distance on the same sample surface (inset of Fig. 3a), and the resulting plot of $R$ versus $X_g$ is fitted to a linear relationship to obtain a fitting parameter $g \equiv \rho_{2D}/\rho_{3D} \times s_{14}$ which measures the ratio of bulk to surface conductivity (see Method) [32]. Figure 3a shows the results of variable probe-spacing spectroscopy for the $x = 0$ compound at $T = 82$ K. The fitted $g = 35.5$, indicating that 97% of the conductivity comes from the bulk channel. The obtained bulk resistivity $\rho_{3D} = 3.04$ mΩ·cm matches well those measured *ex situ* with physical property measurement system (PPMS) on similar samples [6,11]. In contrast, the measurement on $x = 0.64$ compound at $T = 82$ K gives rise to $g = 1.7$, which indicates that the portion of the surface conductivity has increased to 37 % (Fig. 3b). Furthermore, we found $g = 0$ at 10 K, namely 100% of the conduction comes from the surface channel at low temperature. The decrease in temperature is seen to increase $\rho_{3D}$ and decrease $\rho_{2D}$, which is consistent with the expected behaviors of TIs [34]. The results provide direct confirmation on the bulk-insulating behavior of $x = 0.64$ compound and the existence of the topological surface states, corroborating well with the QPI results.

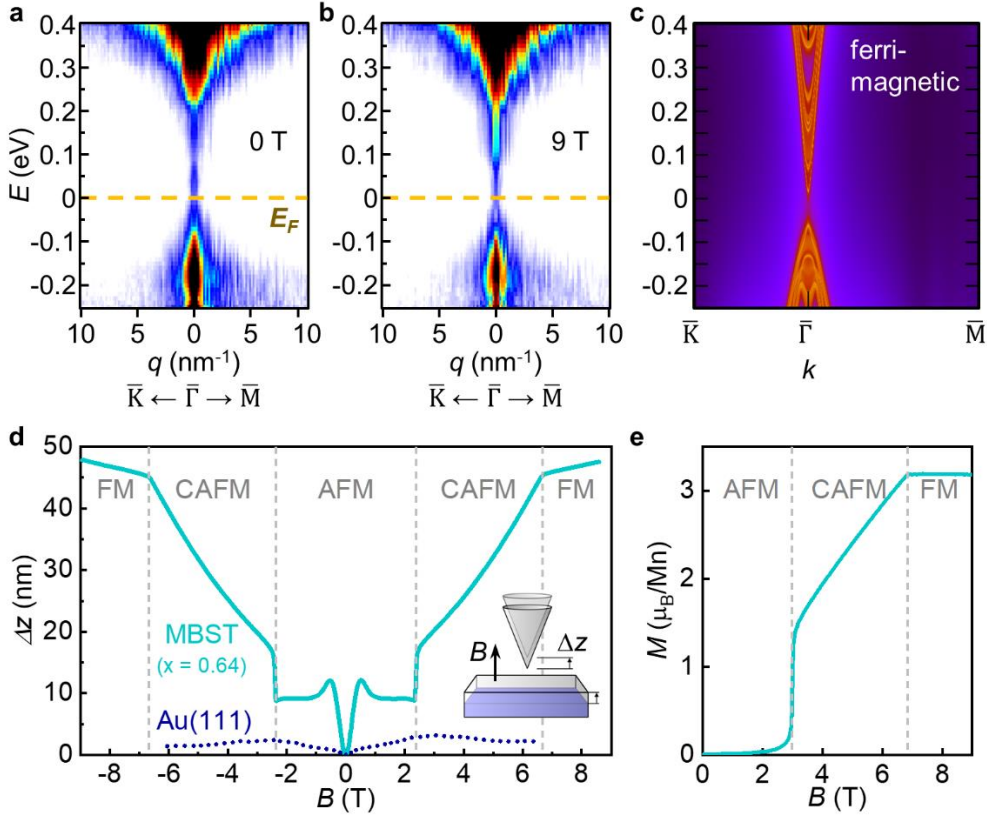

**Fig. 4. The effect of the external magnetic field on MBST. a,b.** QPI measured at 0 T and 9 T, respectively (set point $R_T = 0.3$ GΩ). **c.** DFT calculation of the band structure as projected to (111) surface of $x = 0.64$ compound in the ferrimagnetic state (inset). **d.** Magnetostriction measured with STM. STM was operated in constant current mode with the feedback on, and the tip height was recorded while the out-of-plane magnetic field was varied (lower right inset). **e.** The c-axis magnetization versus magnetic field measured *ex situ* with PPMS for $x = 0.63$ compound (adapted with permission from the ref. [11]. Copyright 2019 by the American Physical Society).

To gain insight into the effect of magnetic structure on the topological phases of MBST, we performed QPI measurements in magnetic field. Applying 9 T out-of-plane magnetic field is expected to change the interlayer AFM order to FM and suppress the anisotropic fluctuations of the local magnetic moments [11]. QPI maps taken at both 0 and 9 T at the same surface location of the $x = 0.64$ compound are comparatively shown in Fig. 4a and b, respectively. The dispersion relations for both the bulk conduction band at $E > 0.2$ eV and valence band at $E < -0.05$ eV are almost identical for both cases, which shows the robustness of the band structure against the magnetic field. The size of the bulk band gap remains unchanged with magnetic field, indicating that the field-induced FM state is an insulator rather than a WSM [3,4]. We note that both Fig. 4a and b show QPI signal minimal at 0~0.05 eV, consistent with the measured 50 meV surface band gap, although the weak QPI signal of the surface states makes it hard to resolve the fine structure around the Dirac point at high magnetic field. In addition, the surface states do not show any change in d$I$/d$V$ spectra as the magnetic field increases from 0 T to 8.5 T for $x = 0$ compound (Fig. S3) [16]. Therefore, both the bulk and surface states are robust against the out-of-plane magnetic field.

The field-induced AFM to FM transition in MBST is further confirmed by a magnetostriction effect detected by STM. Magnetostriction comes from magnetoelastic coupling that causes a change in dimensions in response to varying net magnetization in magnetic materials [35]. To detect the magnetostriction, the STM tip was held in the constant current mode, and the change in tip height $\Delta z$ was recorded while ramping the out-of-plane magnetic field $B$ (inset of Fig.4d). With the increase of $B$, $\Delta z$ shows large jumps at ±2.4 T and clear kinks at ±6.7 T. These transition points match well the behaviors observed in c-axis magnetization measurement with PPMS as shown in Fig. 4e, where magnetic transitions occur sequentially from AFM to CAFM and then to FM states [11]. The change of $\Delta z$ is mainly due to the magnetoelastic coupling in MBST since much smaller variations in Δz and no jumps or kinks are observed by repeating the measurement on a non-magnetic Au(111).

To understand the topological phases of the MBST in the FM state, we employed DFT calculations and considered the two different magnetic configurations: ferrimagnetic and ferromagnetic. Because the measured magnetic moment of Mn in Fig. 4e is less than the theoretical prediction of 4~5$\mu_B$/Mn for Mn$^{2+}$, [4,5,11] we considered the ferrimagnetic case where some Mn ions couple antiferromagnetically with the majority of Mn as shown in Fig. S13. For a fully parallel coupling of all the Mn$^{2+}$ ions (i.e. the FM case), the $x = 0.64$ compound with bulk experimental lattice parameters is a normal semimetal with zero Chern number (Fig. S7). In the ferrimagnetic configuration with only 2/3 of Mn$^{2+}$ ions coupled ferromagnetically, the $x = 0.64$ compound contains a clear gap (Fig. 4c). The finite bulk band gap is similar to the AFM configuration, but has a zero Chern number for $k_z = 0$ and $k_z = 0.5$ (Fig. S8), and the corresponding (111) surface shows a surface band gap.

Furthermore, we compared the topologies of $x = 0.64$ and $x = 0$ compounds in the ferrimagnetic configuration. Ferrimagnetic MnBi$_2$Te$_4$ has band inversion between Bi-p and Te-p states at the Γ point, but its Chern number equals zero for the $k_z = 0$ and $k_z = 0.5$ (Fig. S9). Since the band inversion indicates non-trivial topology, we further calculated the $Z_4$ value using the parity criteria of the occupied bands at the eight time-reversal invariant momenta because MnBi$_2$Te$_4$ contains inversion symmetry [36]. The calculation yields a value of $Z_4 = 2$ (Table S2), which indicates that ferrimagnetic MnBi$_2$Te$_4$ is a topological axion insulator with gapped surfaces on both of the (111) and (110) surfaces (Figs S11, S12). The results suggest that $x = 0.64$ compound in the magnetic field is also most likely an axion insulator since our alloy model constructed using virtual crystal approximation preserves inversion symmetry in MnBi$_2$Te$_4$. We note that this is a surprising result, because by breaking time-reversal symmetry in a typical $Z_2$ insulator, the system should become a semimetal (Weyl or nodal-line semimetal). The fact that the system is an insulator with a zero-Chern number suggests that the $Z_2$ in the AFM state is protected by a combination of time-reversal and translational symmetry. Breaking time-reversal symmetry breaks the translational-symmetry as well, so that the system undergoes a topological phase transition into an axion phase with weaker protection by a non-zero $Z_4$ in the presence of inversion symmetry. As a result, the presence of magnetic field leads to a transition between two non-trivial topological insulating states, which would not require gap closing in the bulk band structures.

In conclusion, the topological phases of MBST with respect to the Sb substitution and external magnetic field are determined based on combined STM measurements and theoretical calculations. The QPI measurement reveals dispersing features from the topological surface states for $x = 0$ and 0.64 compounds inside a bulk band gap, while no such feature is seen in the bulk band gap of $x = 2$ compound. By realizing bulk-insulating state in $x = 0.64$ compound, a surface band gap of 50 meV is detected. Variable probe-spacing transport spectroscopy with multiprobe STM shows surface dominant transport for the $x = 0.64$ compound at 10 K and confirms the existence of the topological surface states. DFT calculations indicate that the $x = 0.64$ compound is a bulk-insulating AFM TI, consistent with the experimental observations. The magnetic transitions from AFM to CAFM and CAFM to FM are observed in magnetostriction measurement with STM. However, the bulk band gap remains unchanged at both AFM and FM states, excluding a possible transition to WSM and suggesting a topological axion insulator in the field-induced FM state. The clarification of the topological phases and relatively large surface band gap in MBST would enable further explorations of emergent quantum properties, such as Majorana fermions at the interface between QAH states and superconductors [37,38], and applications of these materials for topological electronics and computations.

**Methods**

*Scanning tunneling microscopy/spectroscopy:*

STM/STS and QPI maps are acquired by using an Omicron low-temperature 4-probe STM. LEED attached to the preparation chamber is used to check the crystallinity of the surface. Variable probe-spacing measurements are performed with RHK/Unisoku 4-probe STM [31]. STM data in magnetic field are acquired with Unisoku USM1500 with a superconducting coil that generates the magnetic field perpendicular to the surface and maximum strength of 9 T. All STM data are acquired at ultra-high vacuum (UHV) condition ($< 5 \times 10^{-10}$ torr). Measurement temperature is 4.6 K if it not being specified in the main text. STM tips are prepared from etched W wires or mechanically polished PtIr probes from Unisoku, and are further sharpened with focused ion beam (FIB) milling before STM. The metallic behavior of the tips is verified on Au(111) before the measurement. MBST single crystals with various compositions are grown by a flux method as described in references [6,11], and cleaved in the UHV chamber at room temperature. d$I$/d$V$ spectra are acquired by using the lock-in technique with the modulation frequency of 500 Hz and the modulation voltage of 3~10 mV. d$I$/d$V$ spectra in the unit of nS are calibrated by comparing with the numerical derivative of the $I$-$V$ spectra. To generate the QPI maps, d$I$/d$V$ spectra on the regularly spaced grid are measured and then Fourier transformed to visualize the intensity of the scattering wavevector of quasiparticles at a certain energy.

*Variable probe-spacing spectroscopy:*

For the variable probe-spacing transport spectroscopy, four STM tips are placed in a colinear configuration where the two outer probes (tip 1&4) provide source current $I_s$ and the two inner probes (tip 2&3) measures voltage $V_p$ (inset of Fig. 3a). Four probe resistance $R \equiv I_s/V_p$ is measured for the various distance between the inner voltage probes. The sample is modeled as a conductor with two conduction channels, i.e., 3D bulk and 2D surface, that are intertwined at every point on the surface as reported previously [32-34]. $R$ is expressed as

$$R = \frac{V_p}{I_s} = \rho_{2D} \cdot \frac{1}{2\pi} \ln\left[\frac{\left(g+\frac{s_{14}}{s_{12}}\right)\left(g+\frac{s_{14}}{s_{34}}\right)}{\left(g+\frac{s_{14}}{s_{13}}\right)\left(g+\frac{s_{14}}{s_{24}}\right)}\right] = \rho_{2D} \cdot X_g,$$

where $\rho_{2D}$ and $\rho_{3D}$ are resistivities of the surface and bulk, respectively, and $X_g$ is a function of probe distances $s_{ij}$ and bulk-surface conductivity ratio $g \equiv \rho_{2D}/\rho_{3D} \times s_{14}$ [32]. The equation indicates that $R$ scale with $X_g$ linearly if a correct value for $g$ is identified. Thus, we determine $g$ by finding the linear fit of $R$-$X_g$ graph. The values of $\rho_{2D}$ and $\rho_{3D}$ are extracted from the slope of the linear fit and the value of $g$.

*DFT calculation:*

The calculations are done using the PBE+U functional [39,40] in the VASP code with U = 4 eV as in the previous studies[4,41]. The surface projection and topological property of the MnBi$_2$Te$_4$, MnSb$_2$Te4, and MBST ($x = 0.64$) are characterized using the Wannier tight-binding method as implemented in the WannierTools code [42]. Here we utilize atomic like orbitals for Bi's p, Sb's p, Te's p and Mn's d for the Wannier basis obtained from the Wannier90 code [43]. The virtual crystal approximation (VCA) method is used to study the MBST's electronic and topological properties. In our DFT calculations, the experimental lattices for MnBi$_2$Te$_4$ (a = 4.33 Å, c = 40.91 Å)[6], MnSb$_2$Te$_4$ (a = 4.24 Å, c = 40.87 Å) [11], and MBST (x = 0.64, a = 4.30 Å, c = 40.94 Å) [11] are used to perform the relaxation of the internal ions till forces are less than 0.01 eV/ Å with a cutoff energy of 400 eV. The spin-orbit coupling is included in all the electronic and topological characterization. A k-point mesh of 9 × 9 × 3 is used for the AFM configuration, and 9 × 9 × 5 was used for the FM and ferrimagnetic coupling.

**Acknowledgments**

This research was conducted at the Center for Nanophase Materials Sciences, which is a DOE Office of Science User Facility. We acknowledge the assistance of James Burns and Jonathan Poplawsky for focused ion beam milling of the STM tips. The research by J.Y. and S.O. was supported by the U.S. Department of Energy, Office of Science, Basic Energy Sciences, Materials Sciences and Engineering Division. A.D.P. was financially supported by the Oak Ridge National Laboratory's Laboratory Directed Research and Development project (project ID: 7448, PI: P.G.). Part of the research used resources of the National Energy Research Scientific Computing Center (NERSC), a U.S. Department of Energy Office of Science User Facility operated under Contract No. DE-AC02-05CH11231. F.L. acknowledges funding from the Alexander von Humboldt foundation through a Feodor Lynen postdoctoral fellowship.

**Author Contributions**

W.K., M.K., and A.-P.L. designed the experiment. W.K, M.K., M.F., Z.G., and A.-P.L. performed STM measurement. J.Y. grew the single crystal samples. A.D.P., S.O., and P.G performed DFT calculation. W.K. and F.L. performed four-probe transport measurement. A.-

P.L. supervised and coordinated the research. W.K. and A.-P.L. wrote the manuscript with contributions from all authors.

**Competing Interests statement**

The authors declare no competing interests.